# Photonic Exceptional Topological Bound States in the Continuum

Liren Chen[1,*], Jingming Chen[1,2,†,*], Zhen Gao[1,‡]

[1]State Key Laboratory of Optical Fiber and Cable Manufacturing Technology, Department of Electronic and Electrical Engineering, Guangdong Key Laboratory of Integrated Optoelectronics Intellisense, Southern University of Science and Technology, Shenzhen 518055, China

[2]Department of Physics and State Key Laboratory of Optical Quantum Materials, The Hong Kong University of Science and Technology (HKUST), Hong Kong, China

*These authors contribute equally

[†‡]Corresponding author email: gaoz@sustech.edu.cn (Z.G.); chenjingming@ust.hk (J.C.);

**Exceptional bound states in the continuum (EBICs) arise when bound states in the continuum (BICs) with distinct intrinsic losses coalesce at exceptional points (EPs), combining continuum-embedded wave confinement with a non-analytic spectral response. However, conventional EBICs rely on precisely controlled modal coupling and intrinsic loss, making them fragile and difficult to realize experimentally. Here, we introduce bulk-band topology to mitigate this fragility and experimentally realize a photonic exceptional topological BIC (ETBIC) in spoof surface plasmon polariton waveguides by merging two symmetry-protected topological BICs (TBICs) to coalesce at an EP, with the topological design supporting robust localization under symmetry-preserving perturbations. By engineering an intrinsic loss contrast and tuning the intermodal coupling, we directly resolve the coalescence of the complex eigenfrequencies, accompanied by measured near-field distributions consistent with the exceptional eigenstate. Furthermore, weak perturbations induce approximately square-root eigenfrequency splitting, yielding an enhanced spectral response compared with the approximately linear frequency shift of an isolated TBIC. These results establish bulk-band topology as a design principle for exceptional BIC engineering and provide a foundation for EBIC-based photonic sensing.**

## Introduction

Bound states in the continuum (BICs) and exceptional points (EPs) are two distinct singularities in open wave systems, offering complementary mechanisms for controlling wave confinement and spectral response. BICs enable spatially localized waves to coexist with extended states at the same frequency [1]. Their confinement arises from suppressed coupling to the continuum, for example, due to incompatible modal symmetries [2–5]. This decoupling enables strong field confinement and has motivated applications in lasing [6–9], filtering [10,11] and sensing [12–14]. EPs, by contrast, are non-Hermitian degeneracies at which eigenvalues and eigenvectors coalesce [15–19]. Near a second-order EP, suitable weak perturbations produce a square-root eigenvalue splitting, yielding an enhanced spectral response that is unavailable to an isolated nondegenerate mode [20–27].

Recent studies have established exceptional bound states in the continuum (EBICs) by bringing multiple BIC modes to exceptional coalescence [28–30]. These states combine the non-radiative confinement of BICs with the non-analytic eigenfrequency response of EPs, offering a potential platform for sensitive spectral detection. Following theoretical demonstrations in dielectric metasurfaces [28], EBICs have been experimentally observed in terahertz and acoustic systems [31,32]. Nevertheless, their formation requires a delicate balance between modal coupling and intrinsic loss, making their experimental implementation sensitive to parameter variations. Moreover, existing realizations also lack a bulk-band topological origin and the associated protection afforded by bulk–boundary correspondence. Incorporating topological BICs could address this vulnerability by providing robust localized modes as building blocks, offering a systematic framework for designing and realizing robust EBICs.

Topological BICs (TBICs) arise when topological boundary states are embedded in a bulk continuum yet remain symmetry-decoupled from the extended bulk modes [33–46]. In these states, bulk–boundary correspondence supports localization at boundaries, whereas symmetry prevents the localized state from coupling to the continuum. Specifically, in our system, the topological interface state and the coexisting bulk modes possess opposite mirror parities, enforcing modal orthogonality and suppressing their hybridization. The interface state consequently retains its localized character under symmetry-preserving perturbations. These properties make TBICs a promising modal basis for achieving exceptional coalescence while preserving BIC confinement.

Here, we experimentally realize an exceptional topological BIC (ETBIC) in spoof surface plasmon polariton (SSPP) waveguides [47,48] by bringing two TBICs with contrasting intrinsic losses to coalescence at an EP. Each isolated waveguide joins domains with different Zak phases and supports an even-parity interface state embedded in an odd-parity bulk continuum. Spectral and near-field measurements characterize its spectral embedding and spatial localization, while perturbation measurements reveal its persistence when mirror symmetry is preserved and its degradation when that symmetry is broken. We then couple two TBICs with contrasting losses and tune their coupling strength to reach an EP. We directly observe the coalescence of both the eigenfrequencies and eigenstates of the two TBICs, establishing the formation of an ETBIC. Under weak perturbations, the ETBIC exhibits square-root eigenfrequency splitting, whereas an isolated TBIC displays an approximately linear response. These results bring together topological localization, symmetry-protected BIC confinement and EP-enhanced sensitivity in a single photonic system, providing a robust route towards future EBIC-based photonic sensing.

## Results

### Conceptual framework for ETBIC

Figure 1 outlines the construction of the proposed ETBIC, which combines topological localization, symmetry-protected BIC confinement, and non-Hermitian degeneracy. The central panel depicts two TBIC modes with distinct intrinsic loss rates $\gamma_1$ and $\gamma_2$, coupled with strength $\kappa$. The surrounding panels illustrate the spatial localization, continuum embedding, and exceptional-point properties of the resulting hybrid state.

A spatially localized topological interface state forms at the boundary between topologically trivial and nontrivial chains. When its eigenfrequency lies within the continuum of extended bulk states, the interface state becomes spectrally embedded. In the present system, the interface state has even mirror parity, whereas the relevant bulk states have odd parity. These opposite parities enforce orthogonality and suppress hybridization, allowing the interface state to remain localized within the continuum. The resulting symmetry-decoupled topological interface state constitutes a TBIC.

To construct an ETBIC, we introduce reciprocal coupling of strength $\kappa$ between two TBIC modes with a controlled contrast in their intrinsic loss rates. Tuning the coupling brings their complex eigenfrequencies and corresponding eigenstates to coalescence at an EP. The resulting ETBIC combines the topological localization and symmetry-protected decoupling of its constituent TBICs with the non-Hermitian degeneracy of the EP. A weak perturbation $\delta$ near the EP lifts the degeneracy and produces the characteristic square-root eigenfrequency splitting, $\Delta\omega \propto \sqrt{\delta}$.

### Experimental realization and characterization of the TBIC building blocks

To realize a TBIC experimentally, we constructed an SSPP waveguide comprising two topologically distinct domains [Fig. 2(a)]. The unit cells in the two domains differed in the relative displacement between adjacent grooves, while both domains preserved mirror symmetry about the waveguide centerline. Figure 2(b) presents the calculated band structures, which support even- and odd-parity modes. Representative field distributions at $k_x = \pi/a$ confirm the symmetric and antisymmetric profiles of these modes. The corresponding Zak phases identify the left and right unit cells as topologically nontrivial and trivial, respectively. Joining the two domains therefore creates a topological interface that can support a localized state.

Symmetric and antisymmetric excitations were used to selectively excite the even- and odd-parity modes, respectively. Figure 2(c) compares the simulated transmission spectra of both mode families with the measured even-parity spectrum, which agrees closely with the simulation. The even-parity interface resonance at 8.23 GHz (red arrow) lies within the continuum of extended odd-parity bulk modes. Despite this spectral overlap, the opposite mirror parities enforce orthogonality and suppress hybridization between the interface and bulk states. The interface mode thus remains symmetry-decoupled from the bulk continuum in which it is embedded, forming a TBIC.

Near-field measurements further confirmed the spatial localization of the embedded interface mode [Fig. 2(d)]. At 7.06 GHz, corresponding to an even-parity bulk mode, the field extends throughout the waveguide. At 8.23 GHz, the field is concentrated at the interface and decays away from it. Both measured field distributions exhibit the expected even symmetry and agree well with the simulations. Together, spectral embedding and spatial localization provide direct evidence for the TBIC nature of the interface state.

To examine the response to structural disorder, we introduced two types of random perturbations [Fig. 2(e)]. Individual grooves were displaced by $+\delta_i$, $-\delta_i$, or 0, either along the propagation direction (axial perturbations) or transverse to the waveguide (transverse perturbations). Axial perturbations preserve mirror symmetry about the waveguide centerline. Under these perturbations, the TBIC resonance remains well defined, with only a limited frequency shift [left panels of Fig. 2(f)]. This persistence is consistent with the topological origin of the interface state and its continued symmetry-protected decoupling from the odd-parity bulk modes. Transverse perturbations, in contrast, break the protecting mirror symmetry and allow hybridization with the bulk continuum. The resonance consequently broadens and weakens as the perturbation strength increases [right panels of Fig. 2(f)]. These responses indicate robustness to the tested symmetry-preserving disorder and sensitivity to symmetry-breaking perturbations.

We then introduced a tunable intrinsic loss channel to enable subsequent EP realization. A metallic scatterer was placed near the interface, where the TBIC field is strongly localized [Fig. 2(g)]. The scatterer-induced intrinsic loss can be tuned continuously by varying the separation between the scatterer and the SSPP waveguide. At each separation, we adjusted the width of the central strip to compensate for the associated frequency shift, thereby keeping the resonance frequency approximately constant as the loss varied. Decreasing the separation progressively broadened the resonance and reduced its amplitude [Fig. 2(h)], indicating an increased modal decay rate, while hybridization with the odd-parity bulk continuum remained negligible (See Supplementary Information for details) [44]. Combined with frequency compensation, this configuration provides a controllable loss channel for realizing the ETBIC.

**Experimental realization of ETBIC and enhanced eigenfrequency response.**Having established the individual TBIC and a controllable loss channel, we coupled two TBIC modes to realize the ETBIC. The coupled system is described by the effective non-Hermitian Hamiltonian $H = \begin{pmatrix} f_1 - i\gamma_1 & \kappa \\ \kappa & f_2 - i\gamma_2 \end{pmatrix}$, where $f_1$ and $f_2$ are the uncoupled resonance frequencies, $\kappa$ is the reciprocal coupling strength, and $\gamma_1$ and $\gamma_2$ are the intrinsic loss rates. In the experiment, a metallic scatterer introduced additional loss into one waveguide, increasing its total modal loss relative to the other. For frequency-matched TBICs, $f_1 = f_2 = f_0$, the eigenfrequencies are $f_\pm = f_0 - \frac{i(\gamma_1+\gamma_2)}{2} \pm \sqrt{\kappa^2 - \frac{(\gamma_1-\gamma_2)^2}{4}}$. We define the loss contrast as $\gamma = \gamma_2 - \gamma_1 > 0$, with mode 2 designated as the higher-loss mode. The eigenfrequencies coalesce at $\kappa_{\mathrm{EP}} = \gamma/2$, corresponding to a second-order EP [Fig. 3(a)]. For $\kappa > \gamma/2$, the eigenmodes have distinct resonance frequencies and identical decay rates. For $\kappa < \gamma/2$, their resonance frequencies coincide while their decay rates separate. At the EP, both the eigenfrequencies and the corresponding eigenstates coalesce. Under the Hamiltonian convention above, the coalesced eigenstate can be written, up to an arbitrary overall phase, as $|\psi_{\mathrm{EP}}\rangle = \frac{1}{\sqrt{2}}\begin{pmatrix}1\\ -i\end{pmatrix}$. This state has equal modal amplitudes and a relative phase difference of $\pi/2$ between the two TBIC components.

To realize this condition experimentally, we placed two SSPP waveguides supporting nominally identical TBICs in parallel [Fig. 3(b)]. A metallic scatterer introduced additional intrinsic loss into one waveguide, while the coupling strength $\kappa$ was controlled by varying the separation between the waveguides. Simulated and measured near-field distributions near $8.235$ GHz were used to characterize the EP eigenstate [Fig. 3(c)]. The two waveguides exhibited comparable field magnitudes, consistent with nearly equal modal amplitudes. With the overall phase reference fixed, the real and imaginary field components were predominantly localized in opposite waveguides, indicating a relative phase difference close to $\pi/2$. The measured distributions agree well with the simulations and are consistent with the expected coalesced eigenstate.

We further identified the EP by tracking the complex eigenfrequencies as the waveguide separation, and hence $\kappa$, was varied. The simulated and measured transmission spectra [Fig. 3(d)] show two initially separated resonances approaching and merging as $\kappa$ decreases. We extracted the complex eigenfrequencies by fitting these spectra with the model described in the Supplementary Information. Figures 3(e) and 3(f) show the resulting real and imaginary parts, respectively, for both simulations and experiments. In the strong-coupling regime, the eigenmodes have distinct real frequencies and nearly identical decay rates. At the critical coupling, where $\kappa$ equals $\gamma/2$, the two complex eigenfrequencies coalesce. Further reducing the coupling leaves the real parts degenerate while the imaginary parts bifurcate. This simultaneous coalescence, together with the near-field distributions in Fig. 3(c), identifies the EP and supports the formation of the ETBIC.

We next examined the response to a weak perturbation. Starting from $\kappa_{\mathrm{EP}} = \gamma/2$, we introduced a frequency detuning $\delta$ into the lower-loss TBIC. The perturbed Hamiltonian is $H_\delta = \begin{pmatrix} f_0 + \delta - i\gamma_1 & \kappa \\ \kappa & f_0 - i\gamma_2 \end{pmatrix}$. At the EP, the resulting difference between the complex eigenfrequencies is $\Delta f = \sqrt{\delta^2 + 2i\gamma\delta}$. For sufficiently small $\delta$, the leading-order magnitude scales as $|\Delta f| \propto \sqrt{\delta}$. By comparison, the resonance shift of an isolated TBIC varies linearly with the same perturbation, $|\Delta f_{\mathrm{single}}| \propto \delta$. This analysis predicts an enhanced eigenfrequency response for the ETBIC in the weak-perturbation regime.

Experimentally, the detuning was introduced through the axial displacement shown in Fig. 4(a), using the same symmetry-preserving configuration as in the preceding characterization of the isolated TBIC. This perturbation shifted the resonance while preserving the mirror symmetry that decouples the mode from the opposite-parity bulk continuum. For sufficiently small displacements, the induced detuning is approximately proportional to $\delta_x$, as independently indicated by the isolated-TBIC response. The separation between the real parts of the perturbed eigenfrequencies was therefore expected to follow $\Delta f_{\mathrm{EP}} \propto \delta_x^{1/2}$. Figure 4(b) compares representative transmission spectra of an isolated TBIC and the ETBIC before and after perturbation. For the same structural displacement, the ETBIC exhibits a larger frequency splitting than the resonance shift of the isolated TBIC. Figure 4(c) summarizes the extracted responses. For the isolated TBIC, $\Delta f_{\mathrm{single}}$ denotes the resonance-frequency shift relative to the unperturbed state; for the ETBIC, $\Delta f_{\mathrm{EP}}$ denotes the separation between the real parts of the two fitted complex eigenfrequencies. The isolated-TBIC response is approximately linear in the perturbation strength, whereas the ETBIC exhibits a larger splitting across the investigated range. The logarithmic plot in Fig. 4(d) reveals the distinct scaling behaviors. The isolated-TBIC response has a slope close to unity, consistent with $\Delta f_{\text{single}} \propto \delta_x$. The ETBIC response has a slope close to $1/2$, consistent with $\Delta f_{\text{EP}} \propto \delta_x^{1/2}$. This square-root scaling is characteristic of the perturbative response near an EP and supports operation of the coupled TBIC system in its vicinity. Relative to the linear response of an isolated TBIC, the ETBIC therefore exhibits an enhanced eigenfrequency-splitting response to weak perturbations.

## Conclusion

We have experimentally realized a photonic ETBIC in coupled SSPP waveguides by bringing two TBIC modes with contrasting intrinsic losses to an EP. Transmission and near-field measurements establish the spectral embedding and spatial localization of the constituent TBICs, while their perturbation responses reveal robustness to the tested symmetry-preserving disorder and sensitivity to symmetry breaking. Tuning the inter-waveguide coupling produces coalescence of the complex eigenfrequencies, accompanied by near-field distributions consistent with the EP eigenstate. Under weak axial perturbations, the coupled system exhibits an approximately square-root eigenfrequency splitting, compared with the approximately linear resonance shift of an isolated TBIC. These findings establish TBICs as localized modal building blocks for engineering exceptional degeneracies within a bulk continuum. In this construction, bulk-band topology supports interface localization, mirror symmetry suppresses hybridization with the bulk continuum, and the balance between reciprocal coupling and loss contrast sets the EP condition. This approach can, in principle, be extended to higher-order exceptional BICs by coupling additional TBIC modes and appropriately tuning their resonance frequencies, coupling strengths and loss contrasts. The ETBIC architecture thus offers a route to designing photonic systems with tunable exceptional degeneracies supported by topological localization and symmetry-protected BIC confinement..

## Data availability

All data that support the plots within this paper and other findings of this work are available from the corresponding authors upon reasonable request.

## Code availability

All the codes used to generate and/or analyze the data in this work are available from the corresponding authors upon reasonable request.

**Acknowledgments**
Z.G. acknowledges funding from the National Key R&D Program of China (grant no. 2025YFA1412300), National Natural Science Foundation of China (grants no. 62361166627 and 62375118), Guangdong Basic and Applied Basic Research Foundation (grant no. 2024A1515012770), Shenzhen Science and Technology Innovation Commission (grants no. 202308073000209), and High-level Special Funds (grant no. G03034K004).

**Competing Interests**
The authors declare no competing interests.

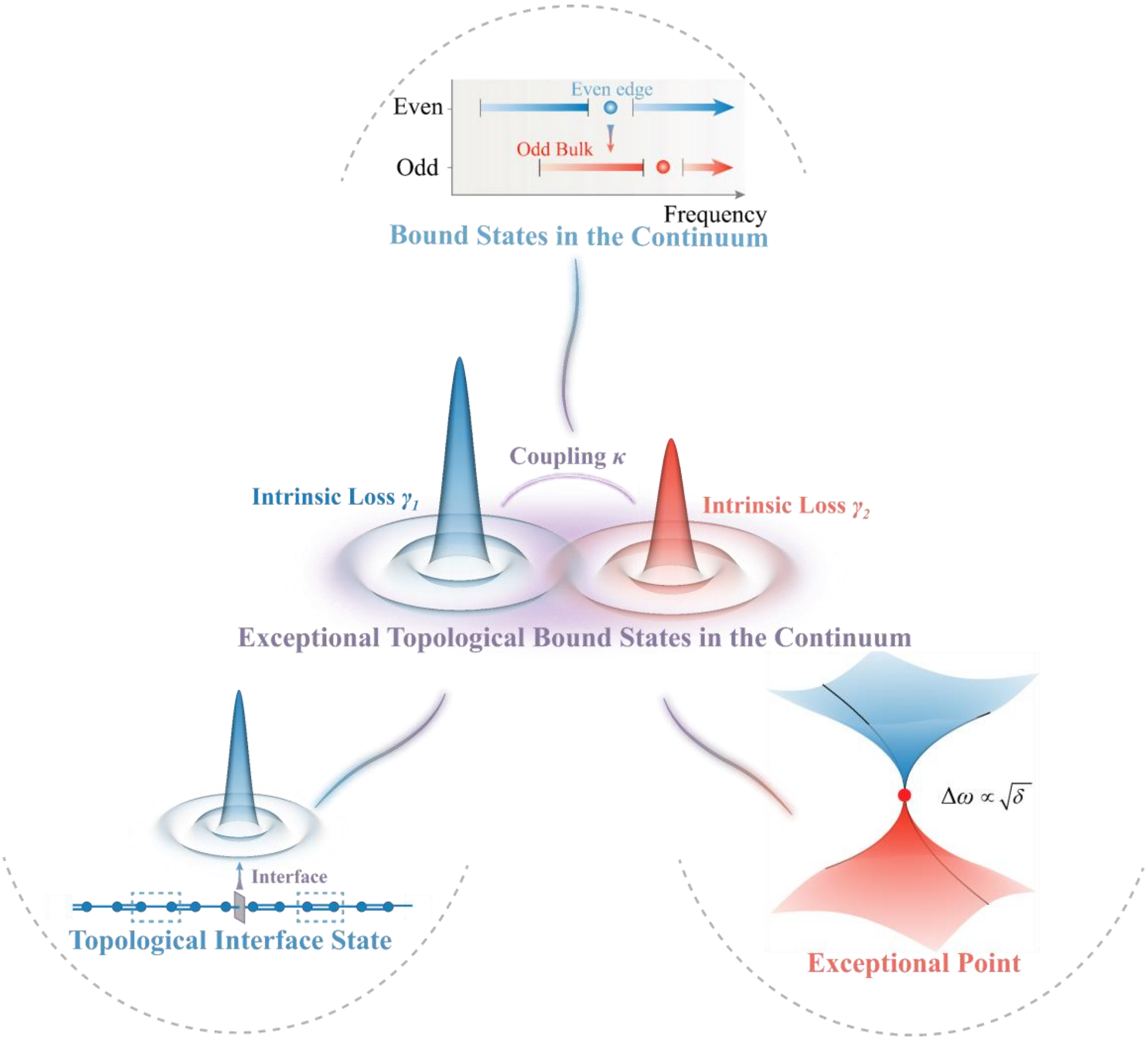

**Fig. 1 | Conceptual framework for the ETBIC.** The central panel depicts two TBIC modes with different intrinsic loss rates $\gamma_1$ and $\gamma_2$, coupled with strength $\kappa$. The surrounding panels illustrate the physical ingredients of the ETBIC. **Lower left:** a topological interface state localized at the boundary between topologically trivial and nontrivial chains. **Upper:** an even-parity interface state embedded in the continuum of odd-parity bulk modes. Opposite mirror parities enforce orthogonality and suppress hybridization, giving rise to a TBIC. **Lower right:** tuning the coupling and loss contrast brings the two TBIC modes to an EP, where their complex eigenfrequencies and eigenstates coalesce. A weak perturbation $\delta$ that lifts the EP degeneracy produces the characteristic square-root splitting, $\Delta\omega \propto \sqrt{\delta}$.

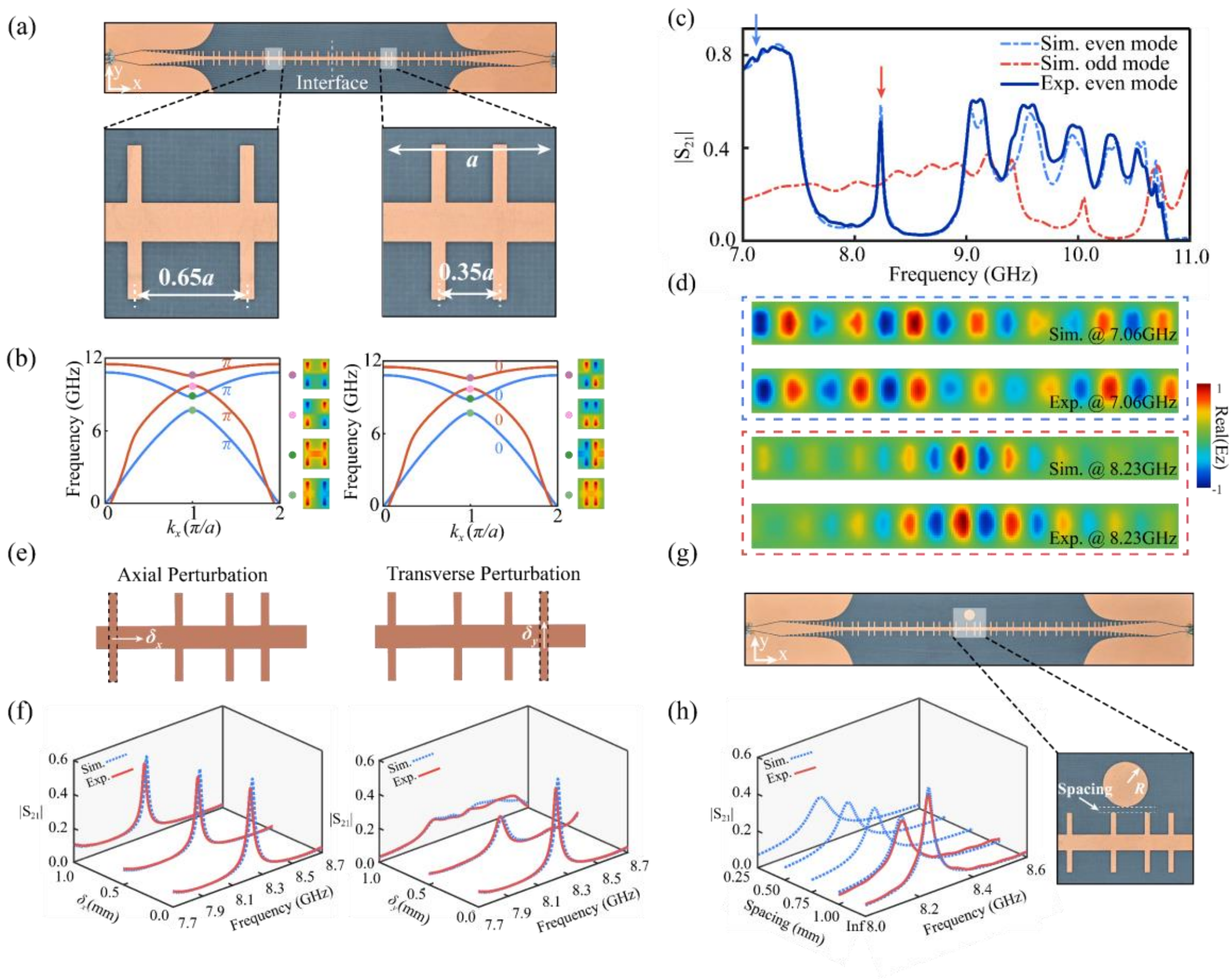

**Fig. 2 | Experimental characterization, perturbation response, and loss control of the TBIC.** **(a)** Photograph of the fabricated SSPP waveguide formed by two topologically distinct chains. Enlarged views show unit cells with groove separations of $0.65a$ (left) and $0.35a$ (right), where $a$ is the lattice period. **(b)** Calculated band structures of the two unit cells. Blue and red bands denote even- and odd-parity modes, respectively; the corresponding Zak phases are indicated beside the bands. The four field profiles beside each diagram correspond to the colored markers at $k_x = \pi/a$. **(c)** Simulated transmission-amplitude spectra of the even-parity (blue dashed) and odd-parity (red dashed) modes, together with the measured even-parity spectrum (dark-blue solid). The interface resonance at 8.23 GHz lies in the even-parity bandgap and overlaps the odd-parity bulk continuum. **(d)** Simulated and measured near-field distributions of the bulk mode at 7.06 GHz (upper pair) and the TBIC at 8.23 GHz (lower pair), corresponding to the blue and red arrows in (c). Within each pair, simulation is shown above measurement. **(e)** Axial and transverse groove displacements, $\delta_x$ and $\delta_y$, which preserve and break mirror symmetry, respectively. **(f)** Transmission spectra for increasing axial (left) and transverse (right) perturbations. Blue dotted and red solid curves denote simulations and measurements, respectively. **(g)** Photograph of the waveguide with a metallic scatterer near the interface; the enlarged view shows the scatterer–waveguide geometry. **(h)** Transmission spectra for decreasing scatterer–waveguide separation, showing progressive broadening and suppression of the TBIC resonance. The line conventions are the same as in (f).

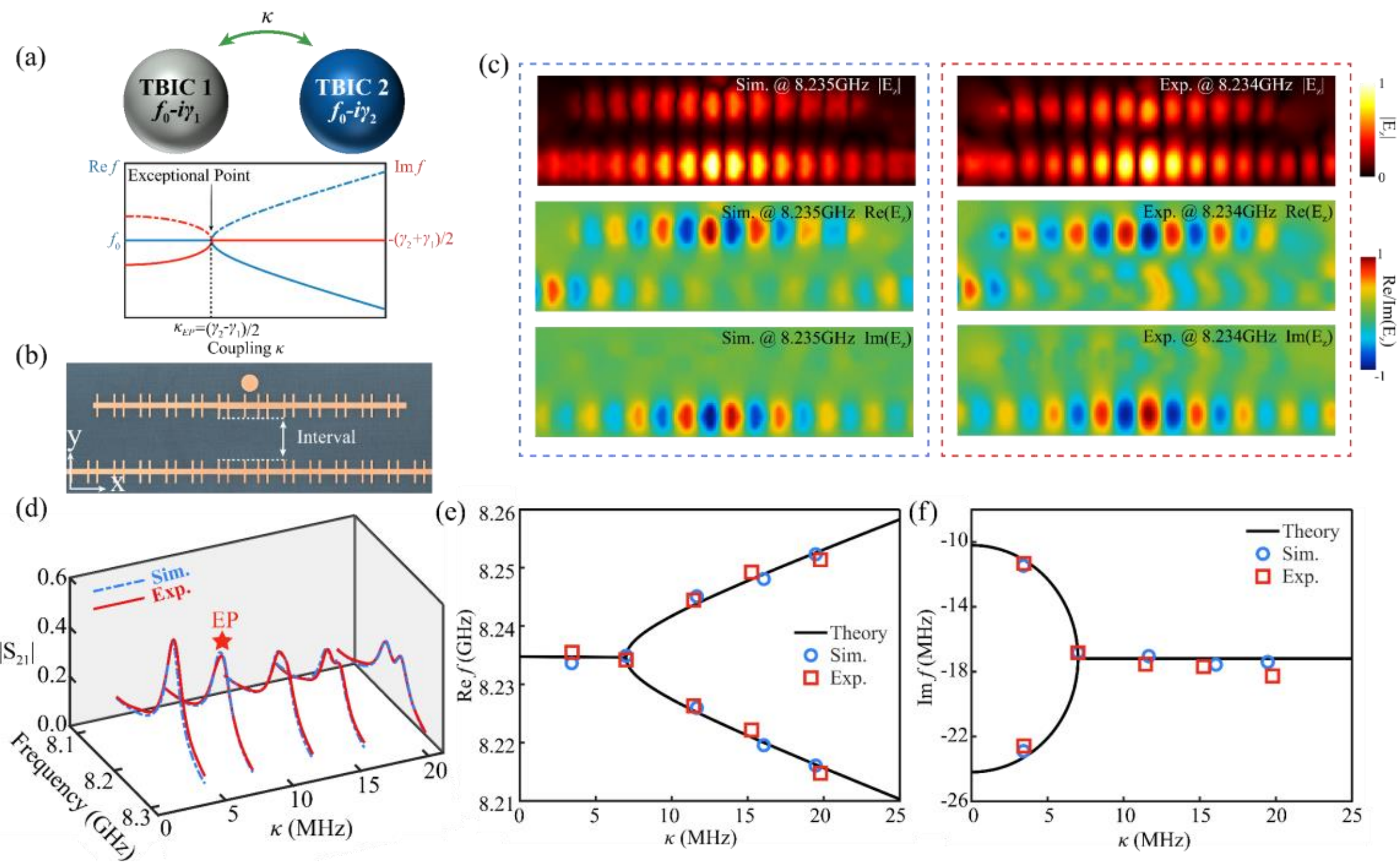

**Fig. 3 | Experimental realization and characterization of the ETBIC. (a)** Upper: two coupled TBIC modes with the same uncoupled resonance frequency $f_0$ and different intrinsic loss rates $\gamma_1$ and $\gamma_2$. Lower: evolution of the real (blue) and imaginary (red) parts of the eigenfrequencies with coupling strength $\kappa$. For a loss contrast $\gamma = \gamma_2 - \gamma_1 > 0$, the EP occurs at $\kappa = \gamma/2$. **(b)** Photograph of the coupled SSPP waveguides. A metallic scatterer introduces additional loss into one waveguide, while their separation controls the coupling strength. **(c)** Near-field distributions from simulations at 8.235 GHz (left) and measurements at 8.234 GHz (right), evaluated at the EP operating point. The rows show the field magnitude, real part, and imaginary part, from top to bottom. **(d)** Simulated (blue dashed) and measured (red solid) transmission-amplitude spectra for several coupling strengths $\kappa$. The red star marks the EP. **(e,f)** Real (e) and imaginary (f) parts of the complex eigenfrequencies as functions of $\kappa$. Blue open circles and red open squares denote values extracted from simulated and measured spectra, respectively. Black solid curves are predictions of the effective Hamiltonian $H$.

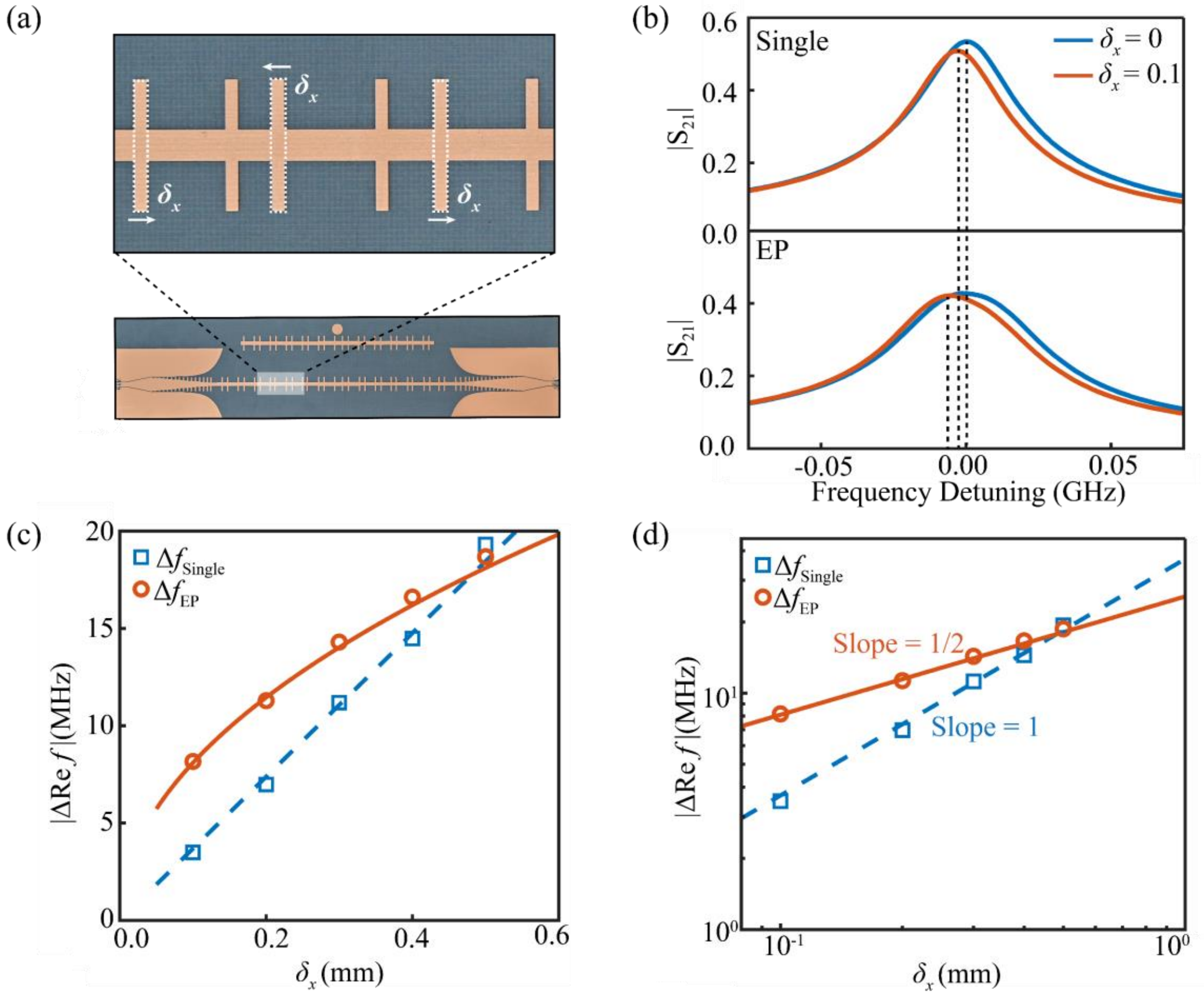

**Fig. 4 | Enhanced eigenfrequency response of the ETBIC to weak perturbations.** **(a)** Photograph of the coupled-waveguide sample and an enlarged view of the symmetry-preserving axial groove displacements of magnitude $\delta_x$ applied to the lower-loss TBIC. **(b)** Measured transmission spectra of an isolated TBIC (upper) and the ETBIC (lower) before perturbation ($\delta_x = 0$, blue) and after perturbation ($\delta_x = 0.1$ mm, orange). **(c)** Frequency responses extracted from spectral fits as functions of $\delta_x$. Blue open squares denote the isolated-TBIC resonance shift $\Delta f_{\text{single}}$ relative to its unperturbed frequency; orange open circles denote the ETBIC splitting $\Delta f_{\text{EP}}$ between the real parts of the two fitted complex eigenfrequencies. The blue dashed and orange solid curves show the corresponding linear and square-root fits. **(d)** The same data on logarithmic axes, with slopes close to 1 (blue dashed) and 1/2 (orange solid), consistent with linear and square-root scaling, respectively.